\documentclass[10pt,twocolumn]{article}

\usepackage[margin=1in]{geometry}
\usepackage{titlesec}
\titleformat{\section}{\large\bfseries}{\thesection.}{0.5em}{}
\titleformat{\subsection}{\normalsize\bfseries}{\thesubsection.}{0.5em}{}

\usepackage{amsmath,amssymb}
\usepackage{graphicx}
\usepackage{booktabs}
\usepackage{tabularray}
\usepackage{arydshln}
\usepackage{caption}
\usepackage{subcaption}
\usepackage{balance}    
\usepackage[colorlinks=true,linkcolor=black,citecolor=black,urlcolor=black]{hyperref}

\title{SpotDiff: Spotting and Disentangling Interference in Feature Space for Subject-Preserving Image Generation}
\author{Yongzhi Li, Saining Zhang, Yibing Chen, Boying Li, Yanxin Zhang, Xiaoyu Du}
\date{}  

\begin{document}
\twocolumn[
\maketitle

\begin{abstract}
Personalized image generation aims to faithfully preserve a reference subject’s identity while adapting to diverse text prompts. Existing optimization-based methods ensure high fidelity but are computationally expensive, while learning-based approaches offer efficiency at the cost of entangled representations influenced by nuisance factors. We introduce SpotDiff, a novel learning-based method that extracts subject-specific features by spotting and disentangling interference. Leveraging a pre-trained CLIP image encoder and specialized expert networks for pose and background, SpotDiff isolates subject identity through orthogonality constraints in the feature space. To enable principled training, we introduce SpotDiff10k, a curated dataset with consistent pose and background variations. Experiments demonstrate that SpotDiff achieves more robust subject preservation and controllable editing than prior methods, while attaining competitive performance with only 10k training samples.
\end{abstract}

\vspace{0.75em}
] 

\begin{figure*}[t]
  \centering
  \includegraphics[width=0.9\linewidth]{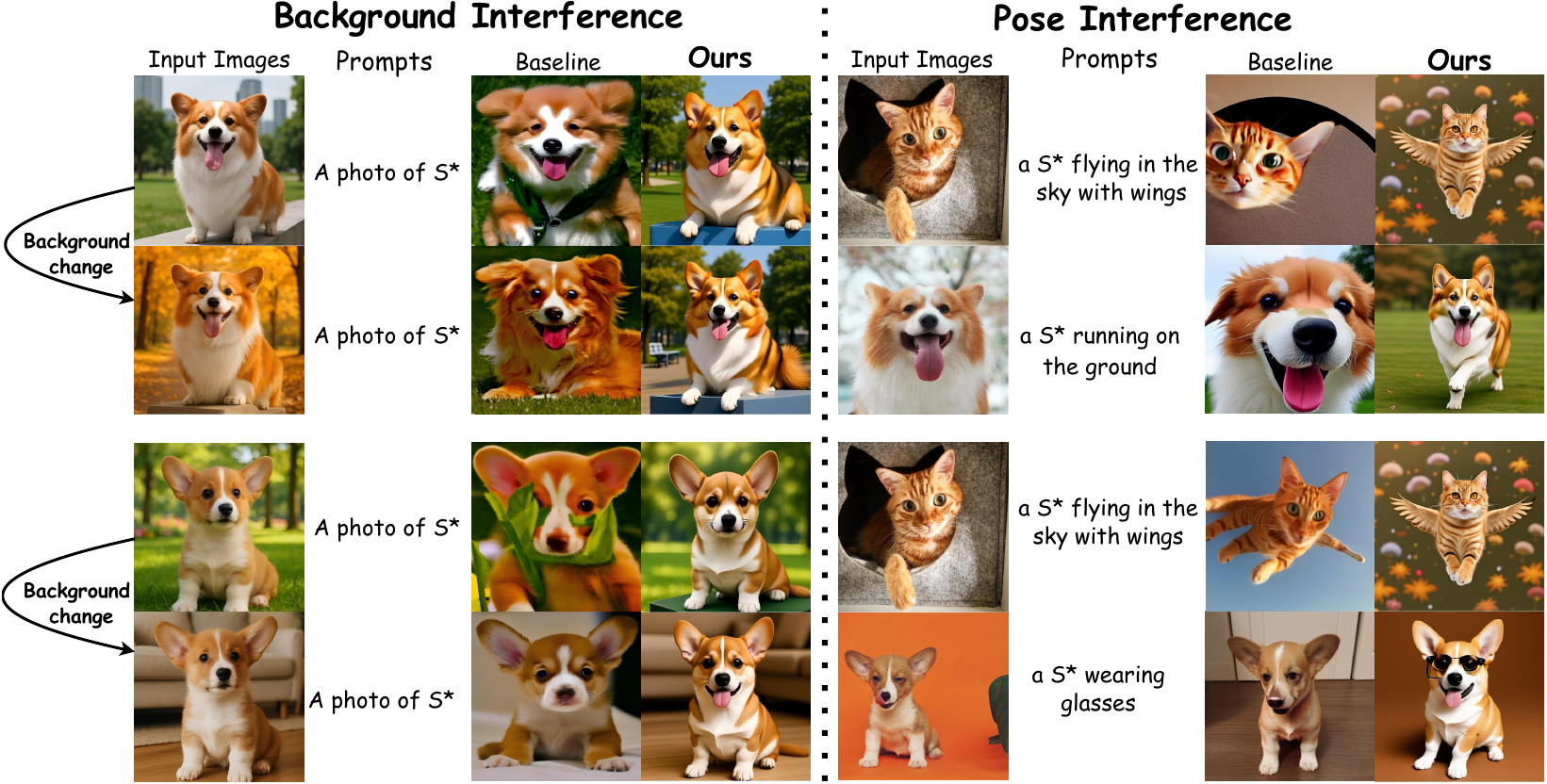}
  \caption{Interference of background and pose. When the background changes, baseline methods exhibit noticeable variations in the generated subject appearance, while our method maintains a consistent subject identity. Furthermore, when editing the subject, baseline methods tend to replicate the pose from the input images, thereby failing to perform accurate subject editing. In contrast, our approach effectively decouples pose from identity, enabling precise subject manipulation.}
  \label{fig1}
\end{figure*}

\section{Introduction}
Diffusion-based text-to-image generative models \cite{ho2020denoising,dhariwal2021diffusion,saharia2022photorealistic,rombach2022high,hoogeboom2023simple} have shown remarkable capabilities, enabling the generation of high-quality and diverse images based on text prompts. Benefiting from the tremendous text-image datasets\cite{schuhmann2022laion}, text-to-image models are able to generate a diverse range of objects, styles, and scenes based on input text prompts, making them highly versatile. However, relying solely on a single text modality is insufficient to generate a personalized concept, as text cannot fully describe it. The methods derived to address the aforementioned issues are referred to as subject-driven generation or personalized generation.

Extensive efforts have been made to achieve personalized image generation, which has demonstrated promising results. Current methods can primarily be categorized into optimization-based methods and learning-based methods. The main idea of optimization-based methods\cite{gal2022image,ruiz2023dreambooth,kumari2023multiconcept,cai2024decoupled} is to treat a rare placeholder $V$ as a learnable object, using 3-5 images containing the target concept to optimize the embedding of $V$ or fine-tune the generative model to link the concept with $V$, which usually requires a noticeable amount of computational resources and time to learn a new concept. To address this, learning-based methods\cite{wei2023elite,li2023blip,song2024moma} are proposed, which try to design encoders to directly map the subject into feature space and combine it with textual information to obtain a multi-modal prompt, thereby eliminating the need for test-time fine-tuning, achieving rapid subject-driven generation. While such methods provide significantly faster inference, they face a major limitation: the encoded representation often becomes entangled with irrelevant attributes in the input image, such as background or pose, which diminishes the precision and controllability of subject-specific generation.

As shown in Figure \ref{fig1}, baseline methods are often adversely affected by irrelevant information. When the background of the input image changes, the baseline models exhibit significant variations in the appearance of the generated subject, such as alterations in the fur color of the Corgi in response to background changes. Additionally, when editing the subject, baseline models tend to preserve the pose from the input image. For instance, when the cat and dog are shown with only their heads in the input, the generated results predominantly depict the head, neglecting the editing prompts. Similarly, for the squinting Corgi at the bottom, the baseline model overemphasizes ``squinting'', ignoring the instructions in the prompt.

Previous research has made significant advancements in decoupling irrelevant information, DETEX \cite{cai2024decoupled} designed two learnable mappers to decouple the pose and background from text embedding mitigating their impact on the subject embedding; Disenbooth \cite{chen2023disenbooth} introduces an Identity-Irrelevant Branch with a learnable mask to separate identity-related features; Song et al. \cite{song2025harmonizing} proposed an Embedding Orchestration method that effectively alleviates pose bias. While these methods improve disentanglement, most of these disentanglement methods are optimization-based, as they are typically tailored to decouple interference factors in a series of images of a specific subject, without systematically investigating how non-subject elements are represented and entangled in the image feature space, making their methods difficult to generalize to unseen subjects.

\begin{figure*}[t]
  \centering
  \includegraphics[width=0.9\linewidth]{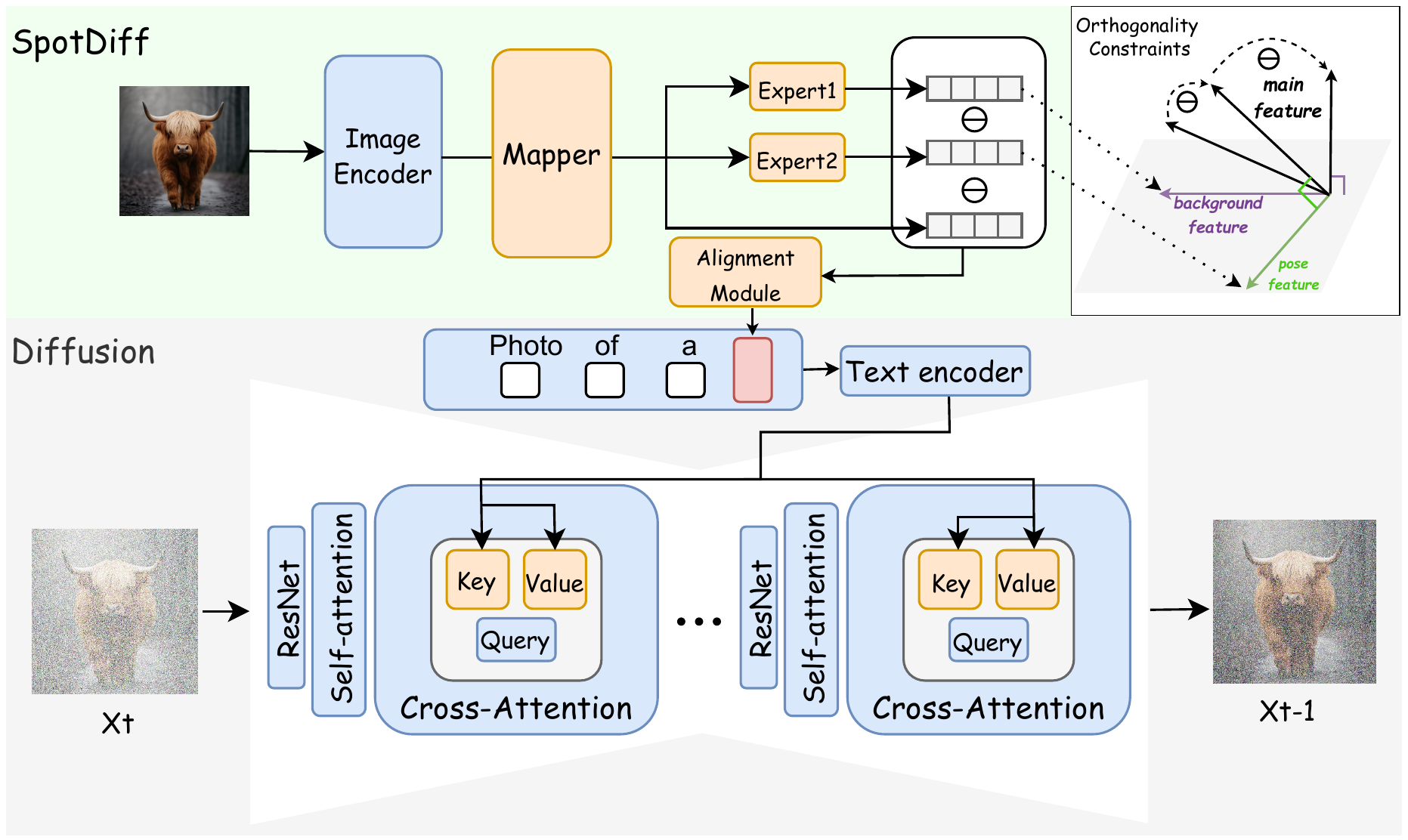}
  \caption{Pipeline Overview: During inference, the image encoder maps the input image to a latent feature space, which is then processed by two expert networks to disentangle relevant semantic features using orthogonality constraints. The resulting feature vectors are aligned and concatenated with a pre-defined text prompt and encoded by the text encoder to form the multi-modal condition guiding the generation process. In the training pipeline, each input image is processed through a noise scheduler to obtain $Z_t$, and at each training step $t$, the model compares the predicted noise to get the gradient. The orange blocks are trainable, while the blue blocks are frozen.}
  \label{fig:s}
\end{figure*}

In this paper, we adopt a new perspective to investigate the manipulability of high-level semantic elements in the feature space. Rather than treating background and pose factors as isolated artifacts to decouple through heuristic designs, we hypothesize that these elements occupy distinct subspaces in the high-dimensional feature space. When the input image is encoded and mapped into feature space, the subject is entangled with unrelated factors like the background, pose, and other attributes. This entanglement leads to ambiguity in generation, as nuisance factors can interfere with the subject-specific features. To address this issue, we build on prior research \cite{shi2024personalized}, which demonstrated the effectiveness of orthogonality constraints in feature decoupling. As shown in the orthogonality constraints part in Figure \ref{fig:s}, by applying these constraints, we ensure that the encoded feature remains orthogonal to both the background and pose features. By enforcing orthogonality, we aim to minimize the influence of irrelevant cues, thereby achieving subject-preserving image generation and editing.

Building on the analysis above, we propose SpotDiff, a novel decoupling model designed to spot and disentangle the interference in the feature space. The framework begins by encoding the input images using CLIP \cite{radford2021CLIP} and a mapper network to obtain the initial main feature vectors. Since the irrelevant information is contained in the initial main feature vectors, it is intuitive to introduce a dedicated network that learns to extract specific nuisance factors from them, thereby assisting feature decoupling. Therefore, the original main feature vectors are passed through specialized expert networks, including the pose expert and background expert, each of which predicts feature vectors corresponding to the specific nuisance factors to be disentangled. Next, we apply orthogonality constraints by subtracting the projections of the main vectors onto the pose and background vectors, ensuring that the main feature becomes orthogonal to both. This process results in new feature vectors that mitigate pose and background interference. Due to the dimensional gap between text and image, an additional alignment module is employed to transform the image features to the appropriate dimension.

To support the training of the aforementioned disentanglement model, we construct SpotDiff10k, a dedicated dataset of 10{,}000 images with controlled variations. Using GPT-4o \cite{openai2024gpt4o} as a generative engine, we synthesize subjects that share the same pose but exhibit different appearances. This allows us to explicitly decouple subject-specific features from pose-related ones. To further control for background interference, we augment each image by replacing its background with diverse scenes. The resulting dataset provides a controlled setting where pose and background can be explicitly separated, thereby supporting the study of feature disentanglement in a principled manner.

\noindent\textbf{The contributions of this work are summarized as follows:}
\begin{itemize}
  \item We propose SpotDiff, a framework that extracts subject-specific features by spotting and disentangling interference. It employs expert networks to predict nuisance features and enforces orthogonality with the original feature to preserve subject-specific information of the input image for controllable image generation.
  \item We construct SpotDiff10k, a novel dataset of 10k images with controlled variations, where subjects share identical poses but differ in appearance, and backgrounds are systematically replaced. This dataset provides a principled foundation for studying the disentanglement of subject, pose, and background in feature space.
  \item Extensive experiments demonstrate that our method not only achieves high-quality image generation but also exhibits strong decoupling ability, effectively mitigating the interference of nuisance factors.
\end{itemize}

\section{Related Work}

\subsection{Text-to-Image Generation}
Diffusion models\cite{sohl2015deep} have recently emerged as the leading paradigm for high-fidelity image synthesis, surpassing the quality and diversity of earlier generative frameworks such as Generative Adversarial Networks (GANs)\cite{goodfellow2020generative}and Variational Autoencoders (VAEs) \cite{kingma2013auto}. These models reverse a noising process, progressively refining a random noise sample into a clear image. For text-to-image generation, diffusion models condition on text descriptions, generating semantically aligned images. Early work adapted DDPMs\cite{ho2020denoising} for conditional generation. More recent models like GLIDE\cite{nichol2021glide} and Imagen\cite{saharia2022photorealistic} further improve realism and semantic alignment.Cross-attention\cite{vaswani2017attention} mechanisms, seen in models like Stable Diffusion \cite{rombach2022high}, improve control over generated content, enabling complex and diverse scenes. Recently, many researchers have explored combining conditioning embeddings from multiple modalities for flexible control, such as integrating text-aligned visual embeddings extracted from images with textual embeddings\cite{pan2023kosmos,sohn2023styledrop,team2024chameleon}. This work also adopts such a multimodal control approach to integrate the various conditioning signals, aiming to leverage the strengths of multimodality.

\subsection{Personalized Generation}
Personalized generation in text-to-image models enables creating customized images based on user-specific data, adapting pre-trained models to reflect individual preferences with minimal extra training. Early personalized generation methods primarily relied on optimization-based approaches\cite{gal2022image,ruiz2023dreambooth,kumari2023multiconcept,hua2023dreamtuner,hao2023vico}, DreamBooth \cite{ruiz2023dreambooth} used a placeholder in the text prompt and fine-tuned all layers to bind a specific subject, preserving its fidelity, while Textual Inversion \cite{gal2022image} achieved a similar result by modifying a trainable pseudo-word to capture both high-level semantics and fine visual details in the embedding space. AttnDreamBooth\cite{pang2024attndreambooth} integrates Textual Inversion and DreamBooth, employing a multi-stage training approach and adding a cross-attention map regularization term to enable personalization. But they all suffer from lengthy test-time fine-tuning. More recent approaches have shifted to learning-based methods\cite{ye2023ip,wei2023elite,li2023blip,shi2024instantbooth,wang2024moa,song2024moma,tan2025ominicontrol,patashnik2025nested}, directly encoding input images to guide the generation. IP-Adapter \cite{ye2023ip} proposes a lightweight, plug-and-play module that equips pre-trained text-to-image diffusion models with the ability to incorporate image prompts alongside text. ELITE \cite{wei2023elite} uses a pretrained CLIP image encoder to extract multi-layer features and applies global and local mapping to project them into the textual space. Instantbooth \cite{shi2024instantbooth} introduces concept and patch encoders to learn image features and uses adapter layers to inject image embeddings, eliminating the need for test-time finetuning. This study builds on learning-based methods to decouple image features, improving the efficiency and quality of personalized generation. However, these methods suffer from two key limitations. First, their visual encoding is susceptible to influence from irrelevant information. Second, they rely on massive datasets—on the scale of 100k samples for general concept learning or 10k for specific class learning—which necessitates substantial computational resources.

\subsection{Feature Disentanglement in Personalized Image Generation}
Personalized text-to-image generation faces the challenge of disentangling a subject's identity from contextual features like pose, background, and style, which can lead to overfitting or identity loss \cite{chen2023disenbooth, song2025harmonizing, hou2025personalized}. Several methods address this by using novel architectures and training strategies. For example, DisenBooth \cite{chen2023disenbooth} learns separate embeddings for identity and irrelevant attributes, while DETEX \cite{cai2024decoupled} uses shared embeddings for the subject and separate ones for pose and background, controlled by mappers and cross-attention loss. Hou et al. \cite{hou2025personalized} apply multi-stage learning to disentangle attributes from a single image. MoA \cite{wang2024moa} and IEDM \cite{chen2025identity} use dual-path systems and a Mixture of Experts module to refine feature fusion. TextBoost \cite{park2024textboost} prevents overfitting with "augmentation tokens" and focuses on fine-tuning the text encoder. Song et al. \cite{song2025harmonizing} address pose-identity entanglement by using orthogonal visual and text embeddings. These methods show that disentangling features is crucial for high-fidelity personalized image generation. Building upon this line of research, our method proposes a novel approach to feature disentanglement in the latent space.  By projecting image features onto specific subspaces for each distinct feature, we efficiently separate a subject's identity from unwanted attributes, giving users precise control over the generated content while maintaining high fidelity.

\section{Preliminaries}
In this section, we first introduce the basic process and condition mechanism of diffusion-based text-image generation.

\noindent \textbf{Diffusion Models:} Our work builds upon Latent Diffusion Models (LDMs) \cite{rombach2022high}, which have become the standard framework for text-to-image generation. Unlike pixel-space diffusion models, LDMs operate in a compressed latent space obtained by a pretrained Variational Autoencoder (VAE). Specifically, an image $x_0$ is first mapped into a latent representation $z_0 = \mathcal{E}(x_0)$ using the VAE encoder $\mathcal{E}$, and after the denoising process, the latent is decoded back into the pixel space $\hat{x}_0 = \mathcal{D}(z_0)$ using the VAE decoder $\mathcal{D}$. This compression significantly reduces computational cost while retaining high-level semantic information.  
Formally, a forward diffusion process is applied to gradually add Gaussian noise to $z_0$:
\begin{equation}
q(z_t \mid z_0) = \mathcal{N}\left(z_t; \sqrt{\alpha_t} z_0, (1 - \alpha_t)\mathbf{I}\right), \quad t = 1, \ldots, T
\end{equation}
where $\{\alpha_t\}_{t=1}^T$ is a predefined noise schedule. The denoising network $\epsilon_\theta$ is trained to predict the added noise given the noisy latent $z_t$ and timestep $t$:
\begin{equation}
\mathcal{L}_{\text{LDM}} = \mathbb{E}_{z_0, t, \epsilon}\left[\left\| \epsilon - \epsilon_\theta(z_t, t, c) \right\|^2 \right]
\label{eq:ldm}
\end{equation}
where $\epsilon \sim \mathcal{N}(0, \mathbf{I})$ and $c$ denotes conditioning information (e.g., text prompts). After training, images are generated by iteratively denoising from Gaussian noise $z_T \sim \mathcal{N}(0, \mathbf{I})$ back to $z_0$, followed by a decoder $\mathcal{D}$ that reconstructs the image $\hat{x}_0 = \mathcal{D}(z_0)$.  

\noindent
\textbf{Condition mechanism:}To enable text-guided generation, LDMs employ a cross-attention mechanism to incorporate textual prompts into the denoising process. Given a latent feature map $h \in \mathbb{R}^{N \times d}$ at a specific layer and a set of textual embeddings $E(c) \in \mathbb{R}^{M \times d}$ obtained from a pretrained text encoder (e.g., CLIP), cross-attention is defined as:
\begin{equation}
\text{Attn}(Q, K, V) = \text{softmax}\left(\frac{QK^\top}{\sqrt{d}}\right) V
\end{equation}

where the queries $Q = hW_Q$, keys $K = E(c)W_K$, and values $V = E(c)W_V$ are linear projections of image and text features. This mechanism allows each spatial location of the latent representation to selectively attend to relevant textual tokens, thereby conditioning the denoising process on semantic guidance from the prompt.

\section{Method}
We present SpotDiff, a framework designed for personalized image generation. The core goal of SpotDiff is to extract subject-specific features by spotting and disentangling interference, including pose and background. By achieving this decoupling, SpotDiff learns robust subject-specific features that can be used to guide text-to-image diffusion models. The overall pipeline of the SpotDiff inference with diffusion process is illustrated in Fig.~\ref{fig:s}. We first fully demonstrate the architecture and the training of our proposed SpotDiff(Sec. 4.1), then we will introduce the pipeline of datasets production(Sec. 4.2).



\subsection{SpotDiff}
In this section, we describe the core components and workflow of the proposed SpotDiff, which is designed to explicitly disentangle the pose and background features within feature space. We first describe the image encoding process, followed by the feature decoupling mechanism. Next, we detail how SpotDiff integrates with the diffusion model, and finally, we present the training objective in mathematical form.

\begin{figure}[h]
  \centering

  \includegraphics[width=\linewidth]{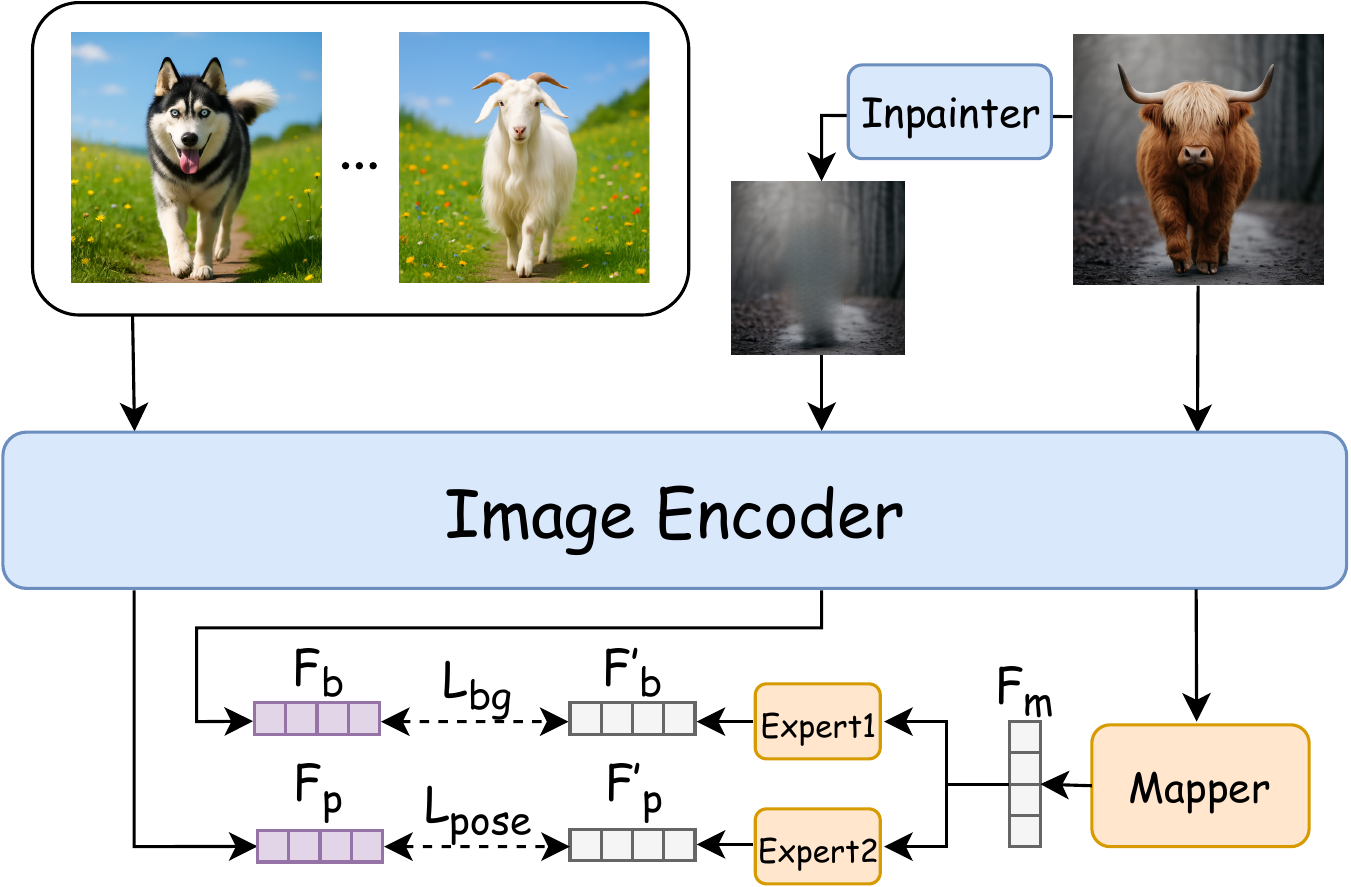}
  \caption{\label{fig:spotdiff1}
SpotDiff framework. The image encoder extracts the feature vectors, which are then passed through expert networks to disentangle pose and background components.}
\end{figure}

\noindent
\textbf{SpotDiff Architecture:}
The proposed SpotDiff begins with encoding a single input image using the CLIP image encoder \cite{radford2021CLIP} and a mapper. To obtain a more comprehensive representation of the image, we follow \cite{wei2023elite}, sampling the features from 5 different layers of CLIP, which generates 5 main feature vectors \( F_{main} \). As shown in Fig.~\ref{fig:spotdiff1}, the feature vectors are then processed by two expert networks: pose and background. Each expert network is responsible for mapping the main feature vector \( F_{main} \) to a corresponding nuisance factor feature. The pose expert focuses on extracting pose-related features $F'_{p}$, and the background expert captures the background features $F'_{b}$. These expert networks learn to specialize in their respective factors by spotting on specific semantic elements. Simultaneously, images with matching pose background are also passed through the CLIP image encoder to obtain ground truth features \( F_b \), and \( F_p \) for background and pose, respectively. These ground truth features are used to train the expert networks to generate accurate representations corresponding to each factor. This architecture allows SpotDiff to effectively decouple the background and pose components, ensuring robust and personalized image generation by guiding the diffusion model through a more precise representation.

\noindent
\textbf{Feature Decoupling by Orthogonality Constraints : }After obtaining the aligned features, we perform orthogonality constraints to extract the subject-specific features. As illustrated in the orthogonality constraints part of Fig.~\ref{fig:s}, the pose and background components are subtracted from the main feature vectors. This operation can be expressed as:

\vspace{-0.25cm}
\begin{equation}
F_{\text{main}} = F \ominus F_{\text{pose}} \ominus F_{\text{background}}
\end{equation}

where the operator \( \ominus \) denotes subtracting the projection of a vector onto another component. For example, \( \mathbf{U} \) represents the vector onto which the projection is calculated, while \( \mathbf{V} \) represents the vector being projected. The projection operation can be written in terms of inner products and norms as:

\vspace{-0.3cm}
\begin{equation}
\mathbf{V} \ominus \mathbf{U} \;=\; \mathbf{V} - \frac{\langle \mathbf{V}, \mathbf{U} \rangle}{\|\mathbf{U}\|^2}\,\mathbf{U}
\end{equation}
\vspace{-0.3cm}

This process effectively makes the main feature orthogonal to interference features, disentangling them from pose and background variations, allowing for more accurate personalization in the generative process.

\noindent
\textbf{Integration with Text and Diffusion Models: } Once the main feature vectors \( F_m \) are obtained, it is processed by an alignment module to match the dimensionality of the text embedding. The aligned vectors are concatenated with a textual prompt to construct a multimodal prompt, which is subsequently encoded by a text encoder. The textual prompt is randomly sampled from 25 CLIP ImageNet\cite{radford2021CLIP} templates. This encoded representation then guides the diffusion model via a cross-attention mechanism to predict noise. Finally, the diffusion model progressively refines the latent representation to generate the final image.

\noindent
\textbf{Training Objective:}  
The training objective is to optimize the expert networks for aligning feature representations between the output of the expert networks and the corresponding ground truth features. As illustrated in ~\ref{fig:spotdiff1}, the expert networks aim to minimize the loss function, which penalizes the misalignment between the predicted and actual feature vectors across different domains, including background and pose. The loss function is defined as:

\begin{equation}
L_{1}
= \frac{1}{N}
  \sum_{i=1}^{N}
  \sum_{k \in \mathcal{K}}
    \left(
        1 -
        \frac{\mathbf{F'}_{k}^{(i)\top} \mathbf{F}_{k}^{(i)}}
             {\|{\mathbf{F'}}_{k}^{(i)}\|_2 \, \|\mathbf{F}_{k}^{(i)}\|_2}
    \right),
\end{equation}

where \( N \) denotes the batch size, and \( \mathcal{K} \) is the set of indices corresponding to different feature categories for each sample. Specifically, \( \mathcal{K} = \{ \text{background}, \text{pose} \} \). The loss terms are computed based on the cosine similarity between the feature vectors produced by the expert networks and the corresponding ground truth features.

In addition to the $L_1$ loss, we also require the LDM loss to train the diffusion model. We adopt the loss function as defined in \ref{eq:ldm}, and following the approach in \cite{wei2023elite}, we add a norm term to obtain the $L_2$ loss:

\vspace{-0.4cm}
\begin{equation}
L_2 = L_{\text{ldm}} + \lambda_1 \|F_{main}\|_1
\end{equation}
\vspace{-0.40cm}

Where $\lambda_1$ is a trade-off hyperparameter. As noted by \cite{kumari2023multiconcept}, cross-attention accounts for the majority of parameter variation during fine-tuning. Therefore, we only optimize the parameters within the cross-attention layers.

Finally, the overall training loss is defined as:

\vspace{-0.45cm}
\begin{equation}
    L = L_2 + \lambda_2 L_1
\end{equation}
\vspace{-0.45cm}

where $\lambda_2$ is a hyperparameter controlling the relative contribution of the $L_1$ loss.

This final training loss balances the contributions from both the LDM loss and the feature alignment loss, ensuring that the model not only learns to generate high-quality images but also maintains the ability to extract subject-specific features by spotting and disentangling interference.

\subsection{SpotDiff10k datasets}
To get the ground truth images matching the pose and background of the main image for the expert network training, we managed to construct new datasets to meet the requirements. To this end, we adopt gpt-4o \cite{openai2024gpt4o} as the image generation backbone, while leveraging Qwen \cite{qwen} to generate description prompts that guide the image synthesis process. The dataset construction procedure is designed to ensure both entity uniqueness and pose consistency, which are essential for producing reliable training data.

The pipeline starts with the generation of unique entity descriptions. We obtain a complete list of entity descriptions in a single request, guaranteeing that each entity is distinct. Next, we use the original image as the reference, replacing the main subject with other entities. To further enforce consistency, detailed pose descriptions are generated by Qwen to precisely capture the posture of the original subject. These pose descriptions are critical for ensuring that all generated samples maintain the same spatial configuration. Finally, we augment the data by changing ten different backgrounds for each subject.

\begin{figure}[h]
  \centering
  \includegraphics[width=\linewidth]{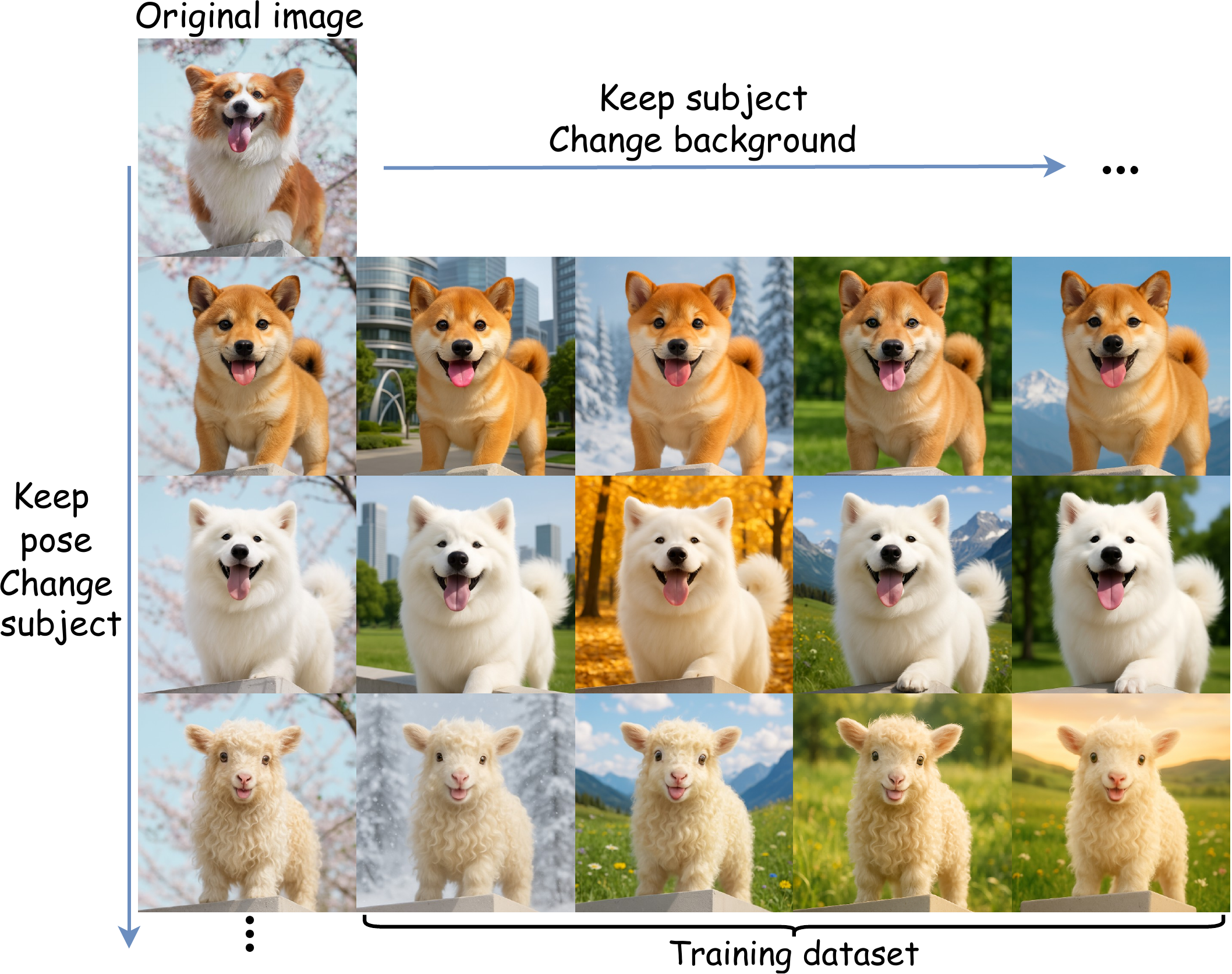}
  \caption{\label{fig:dataset}
           Example from the SpotDiff10k dataset: Given an original image, we generate new subjects with varied appearances while maintaining the same pose. Subsequently, the background of each subject is altered. Each original image is expanded into approximately 100 unique variations.}
\end{figure}

An illustration of this process is shown in Fig.~\ref{fig:dataset}. As demonstrated, given an original image, our method first generates new subjects that share the same pose but differ in appearance, and then alters the background of each subject to create multiple variations. This systematic approach allows each original image to be expanded into approximately 100 unique samples, providing both diversity and structural consistency for training. The original images provided are partly sourced from DreamBooth Benchmark, which are required for evaluation. To avoid data leakage, we strictly separate the provided images from the training set. Moreover, we conduct careful manual inspections to ensure that the newly generated subjects exhibit clear visual distinctions from the provided originals.

\begin{figure*}[tbp]
  \centering
  \includegraphics[width=0.9\linewidth]{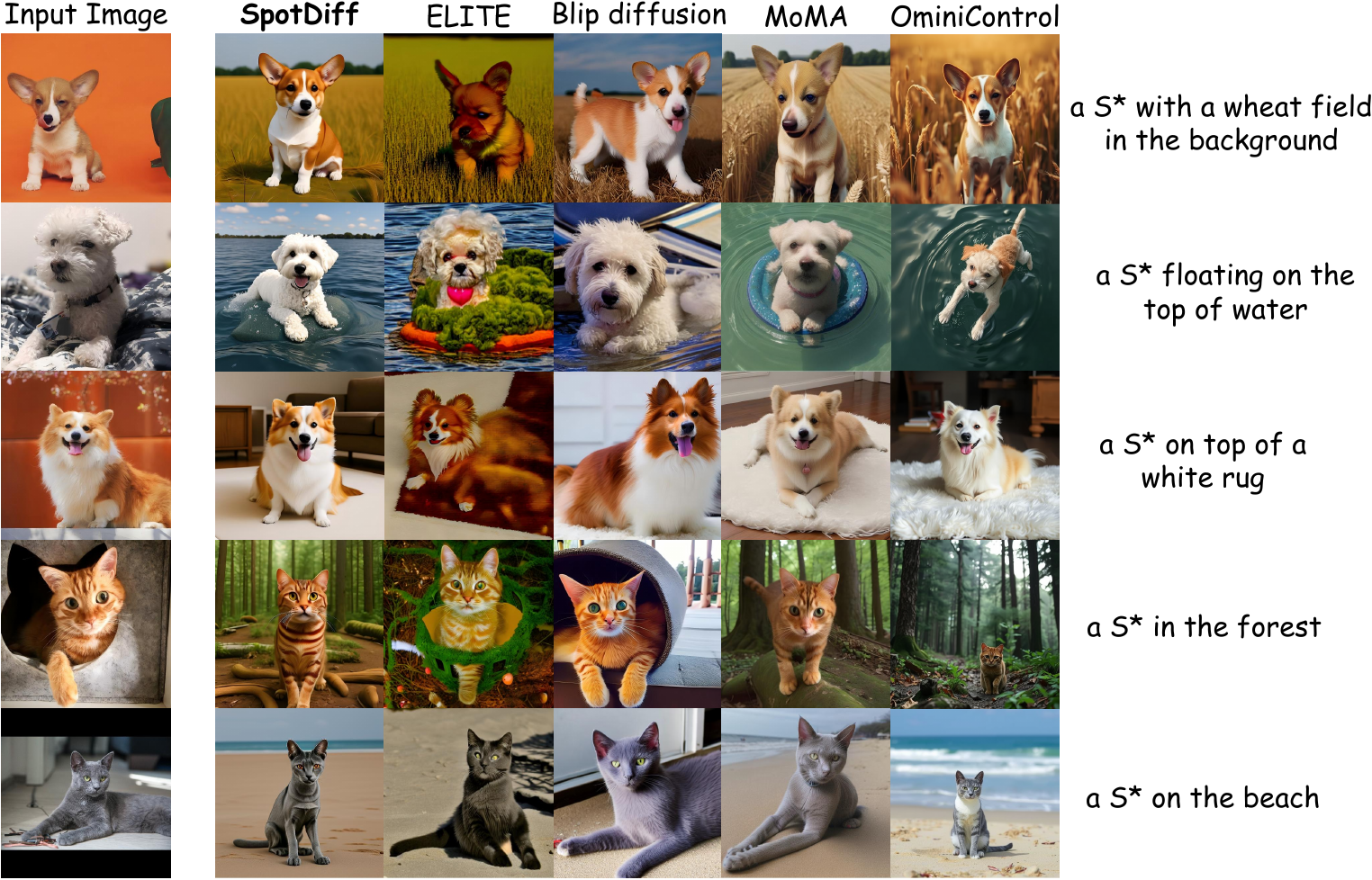}
 \caption{Recontextualization comparison of our method with ELITE, Blip-diffusion, and MoMA. Each row shows the output based on the same input image and prompt.}
    \label{fig:qualitative_comparison}
\end{figure*}

\begin{figure*}[tbp]
    \centering
    \includegraphics[width=\linewidth]{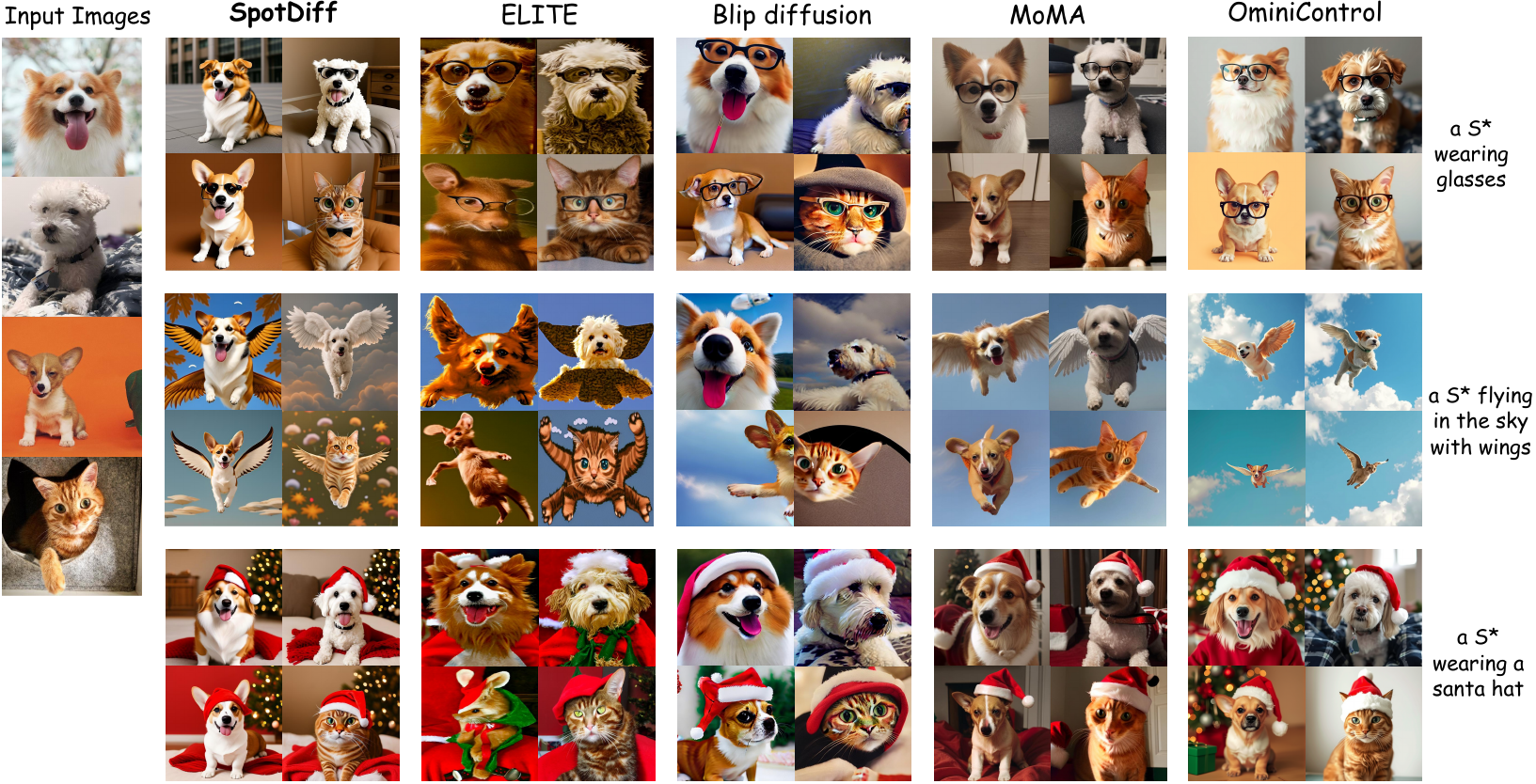}
    \caption{ Accessorization and pose editing results. Each prompt, together with a method, corresponds to an image block; the output images in the block are arranged in order from left to right and from top to bottom, following the vertical sequence of the input images.}
    \label{fig:subject_accessorization}
\end{figure*}

\section{Experiment}

\subsection{Experimental Settings}

\textbf{Datasets:}
We train the entire model on our SpotDiff10k dataset. Each sample in this dataset consists of a main image, which serves as a guide for generation, along with the background segmented using a segmentation model \cite{kirillov2023segment}. The background is further refined through an inpainting method \cite{suvorov2022resolution}. Additionally, several pose-matching images with varying appearances are included, ensuring that the sample design effectively meets the training requirements. For testing, we obtain 30 subjects with obvious pose features from \cite{ruiz2023dreambooth,kumari2023multiconcept}, along with 25 editing prompts from \cite{ruiz2023dreambooth,chen2023disenbooth} for each subject. This results in a total of 750 images for quantitative evaluation. Additionally, for more subject editing visualization results, we construct more editing prompts for subject editing qualitative evaluation.

\textbf{Baseline: }We compare our method against several representative zero-shot subject-driven generation approaches, including ELITE \cite{wei2023elite}, BLIP-Diffusion \cite{li2023blip}, MoMA\cite{song2024moma}, and OminiControl\cite{tan2025ominicontrol}.

\textbf{Evaluation Metric: } 
In this work, we use several metrics to evaluate the performance of the proposed method and baselines. We use the CLIP-T evaluation metric to evaluate the text-image alignment, and to evaluate the fidelity of the generated images, we adopt DINO-I and CLIP-I. For DINO-I, we use the ViTS/16 DINO encoder \cite{caron2021emerging} to encode both the generated and ground truth images, then compute their cosine similarity. Similarly, for CLIP-I, we use the CLIP encoder to obtain embeddings for both the generated and input images and calculate their cosine similarity. For CLIP-T, we compute the cosine similarity between the embeddings of the text prompt and the generated image, with the placeholder S* in the prompt replaced by the subject class. 

\textbf{Implementation Details: }For the diffusion backbone, we adopt Stable Diffusion v2.1~\cite{rombach2022high}. The model is trained using the AdamW optimizer with a learning rate of \(\text{batch\_size} \times 10^{-6}\), where the batch size is set to 8. The loss weights are configured as \(\lambda_1 = 0.01\) and \(\lambda_2 = 0.1\).  All experiments are conducted on a single NVIDIA A800 GPU. During inference, we employ the LMS sampler with 100 timesteps. In the dataset configuration, each sample is equipped with 3 images sharing the same pose. The mapper and expert networks are respectively 5 and 3 layers of MLP with dropout layers, and the alignment module is a 2-layer MLP.

\begin{figure*}[tbp]
    \centering
    \includegraphics[width=0.9\linewidth]{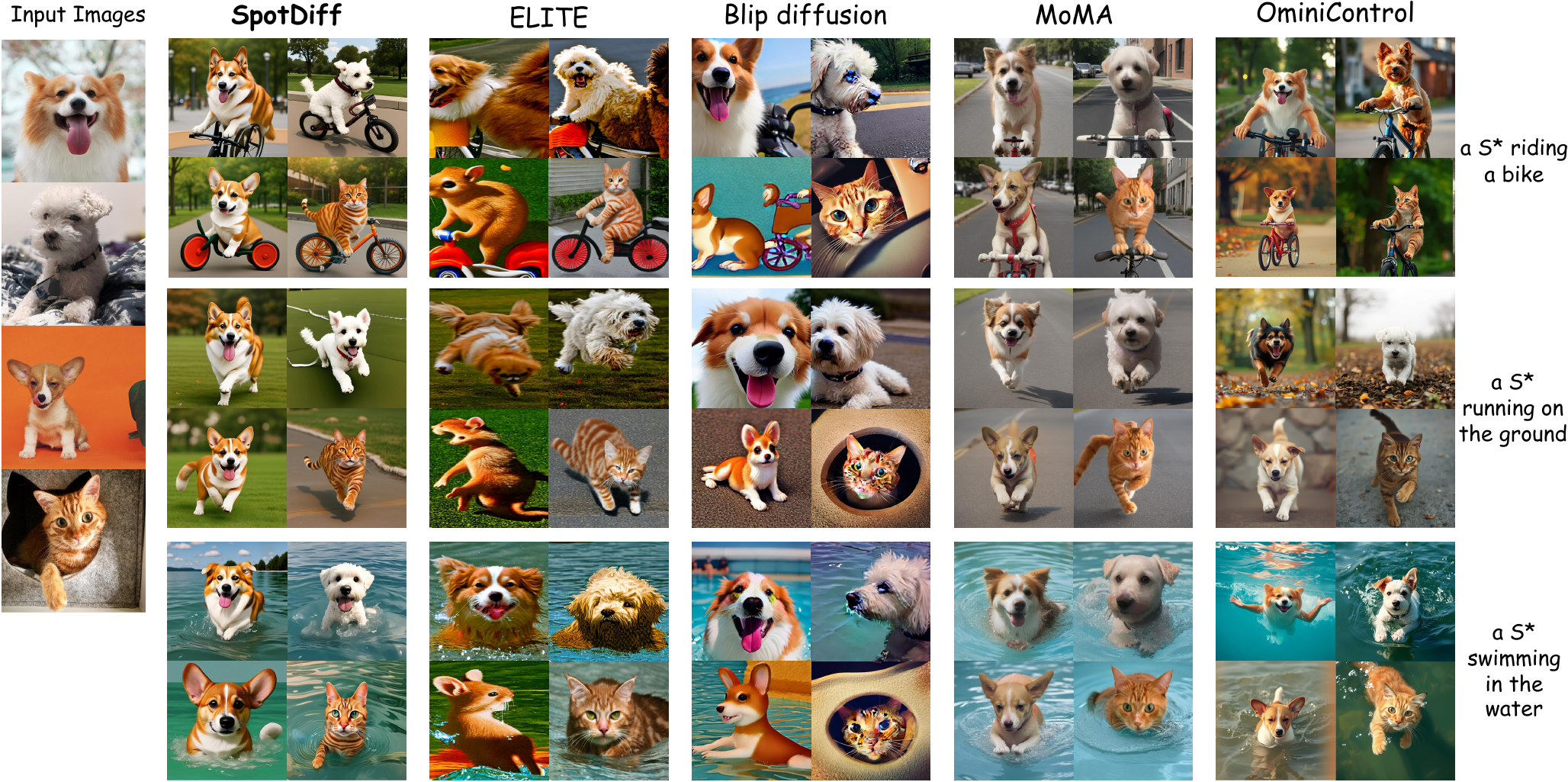}
    \caption{More subject editing visualization results. The same order as ~\ref{fig:pose_editing}, the editability of other models
    is limited by the pose of input images, while ours mitigates the influence.}
    \label{fig:pose_editing}
\end{figure*}

\subsection{Qualitative Results}
We present a qualitative comparison of our method with three other zero-shot approaches: ELITE, Blip-Diffusion, MoMA, and OminiControl, using the same input image and prompt. While Blip-Diffusion, MoMA, and OminiControl require subject class input and do not need the placeholder S*, the S* in their prompts, as shown in the figure, is replaced by the corresponding subject classes in the actual sample process. First, we show the results of recontextualization prompts, as illustrated in Fig.~\ref{fig:qualitative_comparison}. Additionally, we demonstrate the editability of each model using editing prompts, including pose editing and accessorization, as shown in Fig.~\ref{fig:subject_accessorization} and Fig.~\ref{fig:pose_editing}.

\begin{figure*}[tbp]
\centering
\includegraphics[width=0.68\textwidth]{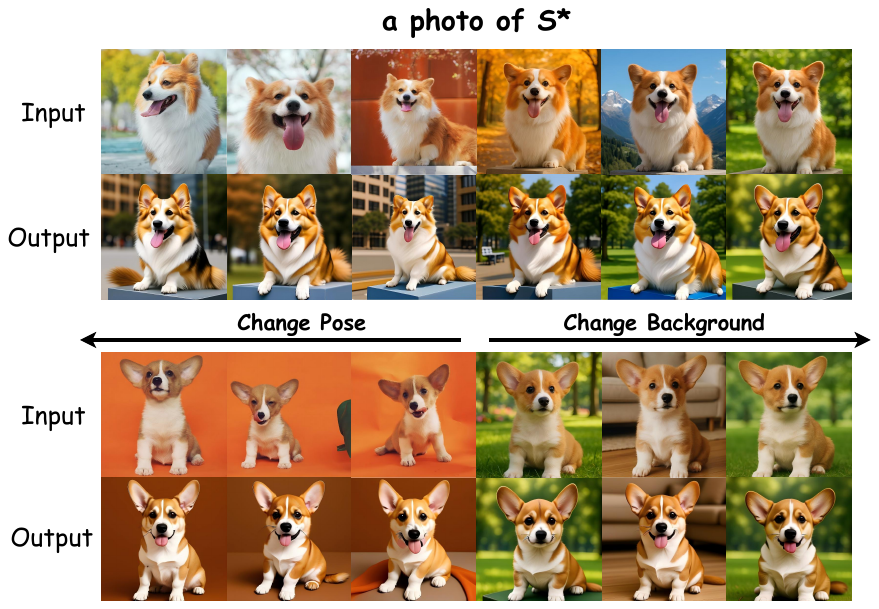}
\caption{Robustness test results showing the performance of our model under varying poses and backgrounds. The left part shows pose variations, while the right part demonstrates background changes. Our method maintains subject fidelity across both types of variations.}
\label{fig:robustness_test}
\end{figure*}

Regarding performance, ELITE shows acceptable text-image alignment but produces visually poor results with noticeable fidelity issues and occasional classification errors. Blip-Diffusion excels in image quality and fidelity but struggles with text-image alignment, leading to some misalignment. MoMA achieves excellent text-image alignment but distorts some details in the generated images. OminiControl demonstrates high generative quality but lacks fidelity to the input image, altering the details of the original subject. In contrast, our method generates high-quality images with strong text-image alignment, maintaining subject consistency and fidelity across a variety of prompts, including recontextualization, pose editing, and accessorization. To further prove our model's resistance to interference from irrelevant information, we conduct additional tests focusing on background and pose variations. As shown in Figure~\ref{fig:robustness_test}, our model consistently produces subject-preserving images well while changing the pose and changing the background, which proves the effectiveness of our proposed decoupling method.

\begin{table}[h]
\centering
\caption{Quantitative Comparison with other zero-shot methods}
\label{tab:zero_shot_comparison}
\begin{tabular}{lccc}
\toprule
Method          & CLIP-T($\uparrow$) & CLIP-I($\uparrow$) & DINO-I($\uparrow$) \\
\midrule
ELITE          & 0.305  & 0.726  & 0.575  \\
Blip-diffusion & 0.286  & 0.803  & 0.631  \\
MoMa           & \textbf{0.335}  & 0.755  & 0.658 \\
OminiControl   & 0.323  & 0.771    &0.643  \\
\hdashline
Ours            & 0.315  & \textbf{0.814}  & \textbf{0.673} \\
\bottomrule
\end{tabular}
\end{table}

\vspace{-0.25cm}

\subsection{Quantitative Evaluation}
As mentioned before, we conduct a quantitative evaluation of our model using three widely adopted metrics: CLIP-T, CLIP-I, and DINO-I. The results show that our model surpasses competing baselines on CLIP-I and DINO-I, demonstrating its strength in generating images with higher fidelity and more faithful visual details. On the other hand, our model does not achieve the best performance on CLIP-T, where MoMA, a model built upon a multimodal large language model (MLLM)\cite{liu2023visual}, outperforms us, which is reasonable, as MLLMs naturally possess strong text-image alignment capabilities, giving MoMA an inherent advantage on this metric.
Furthermore, it is worth emphasizing that our model attains such competitive results using only a dataset of approximately 10k samples, whereas competing methods are typically trained on datasets of around 100k. This comparison highlights the efficiency of our approach and the quality of our dataset.

\begin{table}[h]
    \centering
    \caption{Ablation study. We evaluate the model performance with and without the background expert and pose expert.}
    \label{tab:ablation_study}
    \resizebox{\linewidth}{!}{%
    \begin{tabular}{lccc}
    \toprule
    Method & CLIP-T ($\uparrow$) & CLIP-I ($\uparrow$) & DINO-I ($\uparrow$) \\
    \midrule
    w/o Background Expert & 0.312 & 0.762 & 0.636 \\
    w/o Pose Expert & 0.311 & 0.760 & 0.635 \\
    w/o both & 0.302 & 0.730 & 0.562 \\
    \hdashline
    Ours & \textbf{0.315} & \textbf{0.814} & \textbf{0.673} \\
    \bottomrule
    \end{tabular}%
    }
\end{table}

\subsection{Ablation Study}
To evaluate the impact of each expert module, we perform an ablation study by progressively removing these components. The results, shown in Table~\ref{tab:ablation_study}, demonstrate the importance of each module in improving the overall performance.
When we remove the background expert and pose expert, we observe a decrease in both CLIP-I and DINO-I, suggesting their role in preserving image fidelity. Removing both background and subject experts results in the lowest performance across all metrics.
Our full model, with both experts, achieves the best results. This confirms the effectiveness of each component in enhancing the model's performance.

\section{Conclusion}
In this paper, we propose a novel learning-based subject-driven decoupling method, SpotDiff, to mitigate the interference caused by nuisance factors. Unlike previous decoupling methods, we adopt a new perspective that spots and disentangles interference in the feature space by applying orthogonality constraints. This approach enables high-fidelity, subject-preserving generation and flexible image editing. To support the training of SpotDiff, we construct SpotDiff10k, a specialized dataset containing 10,000 images with controlled variations in pose, subject appearance, and background. This dataset allows us to systematically study the decoupling of identity-related features from other factors.

Our method is not an ad-hoc solution tailored to specific nuisance factors. This general approach suggests potential for decoupling other critical elements, such as style or illumination, which are equally important in personalized image generation. In future work, we aim to explore the decoupling of additional semantic factors, further enhancing the versatility and robustness of our framework.

\balance
\bibliographystyle{unsrt}  

\bibliography{main}       

\begin{thebibliography}{10}

\bibitem{ho2020denoising}
Jonathan Ho, Ajay Jain, and Pieter Abbeel.
\newblock Denoising diffusion probabilistic models.
\newblock {\em Advances in neural information processing systems}, 33:6840--6851, 2020.

\bibitem{dhariwal2021diffusion}
P.~Dhariwal and A.~Nichol.
\newblock {Diffusion Models Beat GANs on Image Synthesis}.
\newblock In {\em Advances in Neural Information Processing Systems}, volume~34, pages 8780--8794, 2021.

\bibitem{saharia2022photorealistic}
Chitwan Saharia, William Chan, Saurabh Saxena, Lala Li, Jay Whang, Emily~L Denton, Kamyar Ghasemipour, Raphael~Gontijo Lopes, Burcu~Karagol Ayan, Tim Salimans, and et~al.
\newblock {Photorealistic Text-to-Image Diffusion Models with Deep Language Understanding}.
\newblock In {\em Advances in Neural Information Processing Systems}, volume~35, pages 36479--36494, 2022.

\bibitem{rombach2022high}
Robin Rombach, Andreas Blattmann, Dominik Lorenz, Patrick Esser, and Bj{\"o}rn Ommer.
\newblock {High-Resolution Image Synthesis with Latent Diffusion Models}.
\newblock In {\em Proceedings of the IEEE/CVF Conference on Computer Vision and Pattern Recognition}, pages 10684--10695, 2022.

\bibitem{hoogeboom2023simple}
Emiel Hoogeboom, Jonathan Heek, and Tim Salimans.
\newblock simple diffusion: End-to-end diffusion for high resolution images.
\newblock In {\em International Conference on Machine Learning}, pages 13213--13232. PMLR, 2023.

\bibitem{schuhmann2022laion}
Christoph Schuhmann and et~al.
\newblock Laion-5b: An open large-scale dataset for training next generation image-text models.
\newblock In {\em Advances in Neural Information Processing Systems}, volume~35, pages 25278--25294, 2022.

\bibitem{gal2022image}
Gal R, Alaluf Y, Atzmon Y, Patashnik O, Bermano AH, Chechik G, and Cohen-Or D.
\newblock An image is worth one word: Personalizing text-to-image generation using textual inversion.
\newblock {\em arXiv preprint arXiv:2208.01618}, 2022.

\bibitem{ruiz2023dreambooth}
Ruiz N, Li~Y, Jampani V, and et~al.
\newblock Dreambooth: Fine tuning text-to-image diffusion models for subject-driven generation.
\newblock In {\em Proceedings of the IEEE/CVF Conference on Computer Vision and Pattern Recognition}, pages 22500--22510, 2023.

\bibitem{kumari2023multiconcept}
Kumari N, Zhang B, Zhang R, Shechtman E, and Zhu JY.
\newblock Multiconcept customization of text-to-image diffusion.
\newblock In {\em Proceedings of the IEEE/CVF Conference on Computer Vision and Pattern Recognition}, pages 1931--1941, 2023.

\bibitem{cai2024decoupled}
Cai Y, Wei Y, Ji~Z, Bai J, Han H, and Zuo W.
\newblock Decoupled textual embeddings for customized image generation.
\newblock In {\em Proceedings of the AAAI Conference on Artificial Intelligence}, pages 909--917, 2024.

\bibitem{wei2023elite}
Wei Y, Zhang Y, Ji~Z, Bai J, Zhang L, and Zuo W.
\newblock Elite: Encoding visual concepts into textual embeddings for customized text-to-image generation.
\newblock In {\em Proceedings of the IEEE/CVF International Conference on Computer Vision}, pages 15943--15953, 2023.

\bibitem{li2023blip}
Dongxu Li, Junnan Li, and Steven Hoi.
\newblock Blip-diffusion: Pre-trained subject representation for controllable text-to-image generation and editing.
\newblock {\em Advances in Neural Information Processing Systems}, 36:30146--30166, 2023.

\bibitem{song2024moma}
Kunpeng Song, Yizhe Zhu, Bingchen Liu, Qing Yan, Ahmed Elgammal, and Xiao Yang.
\newblock Moma: Multimodal llm adapter for fast personalized image generation.
\newblock In {\em European Conference on Computer Vision}, pages 117--132. Springer, 2024.

\bibitem{chen2023disenbooth}
Hong Chen, Yipeng Zhang, Simin Wu, Xin Wang, Xuguang Duan, Yuwei Zhou, and Wenwu Zhu.
\newblock Disenbooth: Identity-preserving disentangled tuning for subject-driven text-to-image generation.
\newblock {\em arXiv preprint arXiv:2305.03374}, 2023.

\bibitem{song2025harmonizing}
Yeji Song, Jimyeong Kim, Wonhark Park, Wonsik Shin, Wonjong Rhee, and Nojun Kwak.
\newblock Harmonizing visual and textual embeddings for zero-shot text-to-image customization.
\newblock In {\em Proceedings of the AAAI Conference on Artificial Intelligence}, volume~39, pages 20549--20557, 2025.

\bibitem{shi2024personalized}
Naichen Shi and Raed Al~Kontar.
\newblock Personalized pca: Decoupling shared and unique features.
\newblock {\em Journal of machine learning research}, 25(41):1--82, 2024.

\bibitem{radford2021CLIP}
Alec Radford, Jong~Wook Kim, Chris Hallacy, Aditya Ramesh, Gabriel Goh, Sandhini Agarwal, Girish Sastry, Amanda Askell, Pamela Mishkin, Jack Clark, et~al.
\newblock Learning transferable visual models from natural language supervision.
\newblock In {\em International conference on machine learning}, pages 8748--8763. PmLR, 2021.

\bibitem{openai2024gpt4o}
OpenAI.
\newblock Gpt-4o.
\newblock \url{https://openai.com/research/gpt-4o}, 2024.
\newblock Large language model.

\bibitem{sohl2015deep}
Jascha Sohl-Dickstein, Eric Weiss, Niru Maheswaranathan, and Surya Ganguli.
\newblock Deep unsupervised learning using nonequilibrium thermodynamics.
\newblock In {\em International conference on machine learning}, pages 2256--2265. pmlr, 2015.

\bibitem{goodfellow2020generative}
Ian Goodfellow, Jean Pouget-Abadie, Mehdi Mirza, Bing Xu, David Warde-Farley, Sherjil Ozair, Aaron Courville, and Yoshua Bengio.
\newblock Generative adversarial networks.
\newblock {\em Communications of the ACM}, 63(11):139--144, 2020.

\bibitem{kingma2013auto}
Diederik~P Kingma and Max Welling.
\newblock Auto-encoding variational bayes.
\newblock {\em arXiv preprint arXiv:1312.6114}, 2013.

\bibitem{nichol2021glide}
Alex Nichol, Prafulla Dhariwal, Aditya Ramesh, Pranav Shyam, Pamela Mishkin, Bob McGrew, Ilya Sutskever, and Mark Chen.
\newblock Glide: Towards photorealistic image generation and editing with text-guided diffusion models.
\newblock {\em arXiv preprint arXiv:2112.10741}, 2021.

\bibitem{vaswani2017attention}
Ashish Vaswani, Noam Shazeer, Niki Parmar, Jakob Uszkoreit, Llion Jones, Aidan~N Gomez, {\L}ukasz Kaiser, and Illia Polosukhin.
\newblock Attention is all you need.
\newblock {\em Advances in neural information processing systems}, 30, 2017.

\bibitem{pan2023kosmos}
Xichen Pan, Li~Dong, Shaohan Huang, Zhiliang Peng, Wenhu Chen, and Furu Wei.
\newblock Kosmos-g: Generating images in context with multimodal large language models.
\newblock {\em arXiv preprint arXiv:2310.02992}, 2023.

\bibitem{sohn2023styledrop}
Kihyuk Sohn, Nataniel Ruiz, Kimin Lee, Daniel~Castro Chin, Irina Blok, Huiwen Chang, Jarred Barber, Lu~Jiang, Glenn Entis, Yuanzhen Li, et~al.
\newblock Styledrop: Text-to-image generation in any style.
\newblock {\em arXiv preprint arXiv:2306.00983}, 2023.

\bibitem{team2024chameleon}
Chameleon Team.
\newblock Chameleon: Mixed-modal early-fusion foundation models, 2024.
\newblock {\em URL https://arxiv. org/abs/2405.09818}, 9(8), 2024.

\bibitem{hua2023dreamtuner}
Miao Hua, Jiawei Liu, Fei Ding, Wei Liu, Jie Wu, and Qian He.
\newblock Dreamtuner: Single image is enough for subject-driven generation.
\newblock {\em arXiv preprint arXiv:2312.13691}, 2023.

\bibitem{hao2023vico}
Shaozhe Hao, Kai Han, Shihao Zhao, and Kwan-Yee~K Wong.
\newblock Vico: Plug-and-play visual condition for personalized text-to-image generation.
\newblock {\em arXiv preprint arXiv:2306.00971}, 2023.

\bibitem{pang2024attndreambooth}
Lianyu Pang, Jian Yin, Baoquan Zhao, Feize Wu, Fu~Lee Wang, Qing Li, and Xudong Mao.
\newblock Attndreambooth: Towards text-aligned personalized text-to-image generation.
\newblock {\em Advances in Neural Information Processing Systems}, 37:39869--39900, 2024.

\bibitem{ye2023ip}
Ye~H, Zhang J, Liu S, Han X, and Yang W.
\newblock Ip-adapter: Text compatible image prompt adapter for text-to-image diffusion models.
\newblock {\em arXiv preprint arXiv:2308.06721}, 2023.

\bibitem{shi2024instantbooth}
Shi J, Xiong W, Lin Z, and Jung HJ.
\newblock Instantbooth: Personalized text-to-image generation without test-time finetuning.
\newblock In {\em Proceedings of the IEEE/CVF Conference on Computer Vision and Pattern Recognition}, pages 8543--8552, 2024.

\bibitem{wang2024moa}
Kuan-Chieh Wang, Daniil Ostashev, Yuwei Fang, Sergey Tulyakov, and Kfir Aberman.
\newblock Moa: Mixture-of-attention for subject-context disentanglement in personalized image generation.
\newblock In {\em SIGGRAPH Asia 2024 Conference Papers}, pages 1--12, 2024.

\bibitem{tan2025ominicontrol}
Zhenxiong Tan, Songhua Liu, Xingyi Yang, Qiaochu Xue, and Xinchao Wang.
\newblock Ominicontrol: Minimal and universal control for diffusion transformer.
\newblock 2025.

\bibitem{patashnik2025nested}
Or~Patashnik, Rinon Gal, Daniil Ostashev, Sergey Tulyakov, Kfir Aberman, and Daniel Cohen-Or.
\newblock Nested attention: Semantic-aware attention values for concept personalization.
\newblock In {\em Proceedings of the Special Interest Group on Computer Graphics and Interactive Techniques Conference Conference Papers}, SIGGRAPH Conference Papers '25. Association for Computing Machinery, 2025.

\bibitem{hou2025personalized}
Xiangyu Hou.
\newblock Personalized text-to-image generation with attribute disentanglement and feature embedding.
\newblock In {\em Computer Science Undergradaute Conference 2025@ XJTU}.

\bibitem{chen2025identity}
Kewen Chen, Xiaobin Hu, and Wenqi Ren.
\newblock Identity-preserving text-to-image generation via dual-level feature decoupling and expert-guided fusion.
\newblock {\em arXiv preprint arXiv:2505.22360}, 2025.

\bibitem{park2024textboost}
NaHyeon Park, Kunhee Kim, and Hyunjung Shim.
\newblock Textboost: Towards one-shot personalization of text-to-image models via fine-tuning text encoder.
\newblock {\em arXiv preprint arXiv:2409.08248}, 2024.

\bibitem{qwen}
Jinze Bai, Shuai Bai, Yunfei Chu, Zeyu Cui, Kai Dang, Xiaodong Deng, Yang Fan, Wenbin Ge, Yu~Han, Fei Huang, Binyuan Hui, Luo Ji, Mei Li, Junyang Lin, Runji Lin, Dayiheng Liu, Gao Liu, Chengqiang Lu, Keming Lu, Jianxin Ma, Rui Men, Xingzhang Ren, Xuancheng Ren, Chuanqi Tan, Sinan Tan, Jianhong Tu, Peng Wang, Shijie Wang, Wei Wang, Shengguang Wu, Benfeng Xu, Jin Xu, An~Yang, Hao Yang, Jian Yang, Shusheng Yang, Yang Yao, Bowen Yu, Hongyi Yuan, Zheng Yuan, Jianwei Zhang, Xingxuan Zhang, Yichang Zhang, Zhenru Zhang, Chang Zhou, Jingren Zhou, Xiaohuan Zhou, and Tianhang Zhu.
\newblock Qwen technical report.
\newblock {\em arXiv preprint arXiv:2309.16609}, 2023.

\bibitem{kirillov2023segment}
Alexander Kirillov, Eric Mintun, Nikhila Ravi, Hanzi Mao, Chloe Rolland, Laura Gustafson, Tete Xiao, Spencer Whitehead, Alexander~C Berg, Wan-Yen Lo, et~al.
\newblock Segment anything.
\newblock In {\em Proceedings of the IEEE/CVF international conference on computer vision}, pages 4015--4026, 2023.

\bibitem{suvorov2022resolution}
Roman Suvorov, Elizaveta Logacheva, Anton Mashikhin, Anastasia Remizova, Arsenii Ashukha, Aleksei Silvestrov, Naejin Kong, Harshith Goka, Kiwoong Park, and Victor Lempitsky.
\newblock Resolution-robust large mask inpainting with fourier convolutions.
\newblock In {\em Proceedings of the IEEE/CVF winter conference on applications of computer vision}, pages 2149--2159, 2022.

\bibitem{caron2021emerging}
Mathilde Caron, Hugo Touvron, Ishan Misra, Herv{\'e} J{\'e}gou, Julien Mairal, Piotr Bojanowski, and Armand Joulin.
\newblock Emerging properties in self-supervised vision transformers.
\newblock In {\em Proceedings of the IEEE/CVF international conference on computer vision}, pages 9650--9660, 2021.

\bibitem{liu2023visual}
Haotian Liu, Chunyuan Li, Qingyang Wu, and Yong~Jae Lee.
\newblock Visual instruction tuning.
\newblock {\em Advances in neural information processing systems}, 36:34892--34916, 2023.

\end{thebibliography}


\newpage
\end{document}